\documentclass[11pt]{article}
\pdfoutput=1 
\usepackage{graphicx}
\usepackage{jheppub} 
\usepackage{enumitem}
\usepackage[T1]{fontenc} 
\usepackage{multirow}    
\usepackage{comment}
\usepackage{amsmath}
\usepackage{amsfonts}
\usepackage{amssymb}
\usepackage{breqn}
\usepackage{bbm}
\usepackage{graphicx}
\usepackage{caption}
\usepackage{slashed}

\def\beq{\begin{equation}}
\def\eeq{\end{equation}}

\newcommand{\bea}{\begin{eqnarray}\begin{aligned}}
\newcommand{\eea}{\end{aligned}\end{eqnarray}}
\newcommand{\bmtx}{\begin{pmatrix}}
\newcommand{\emtx}{\end{pmatrix}}

\newcommand{\neff}{\Delta N_{\rm{eff}}}
\newcommand{\tev}{\text{TeV}}
\newcommand{\gev}{\text{GeV}}

\newcommand{\ztwo}{\ensuremath{Z_2}}
\newcommand{\vev}[1]{{\langle #1 \rangle}}
\newcommand{\eff}{\rm{eff}}
\title{A Portalino to the Twin Sector}

\author{Di Liu and}
\author{Neal Weiner}
\affiliation{Center for Cosmology and Particle Physics, Department of Physics, New York University, New York, NY 10003, USA}

\emailAdd{diliu@nyu.edu}
\emailAdd{neal.weiner@nyu.edu}
\abstract{Extensions of the Standard Model are often highly constrained by cosmology.  The presence of new states can dramatically alter observed properties of the universe by the presence of additional matter or entropy. In particular, attempts too solve the hierarchy problem through naturalness invariably predict new particles near the weak scale which come into thermal equilibrium. Without a means to deposit this energy into the SM,  these models are often excluded. Scenarios of ``neutral naturalness'' especially, such as the Twin Higgs frequently suffer from this. 

However, the Portalino, a singlet fermion that marries gauge neutral fermion operators, can naturally help provide a portal for entropy to return to the SM and to lift fermionic degrees of freedom in the Twin Sector. Together with spontaneous breaking of the $Z_2$  ${\rm SM \leftrightarrow {\rm Twin}}$ symmetry, there are new opportunities to confront the cosmological challenges of these models. 

Here, we attempt to develop such ideas. We shall show how one can lift many of the light fields by breaking $\ztwo$ with a $U(1)_Y$ scalar and its Twin partner. The introduction of Portalinos can lift the remaining degrees of freedom. We shall find that such models are highly constrained by precision SM measurements, motivating moderate extensions beyond this. We will discuss two, one with  additional weak matter and another with additional colored matter. The weak model will involve simple two Higgs doublet models on top of $\ztwo$ breaking. The strong model will involve the presence of new leptoquarks and diquarks. We will discuss the implications for neutrino masses from radiative corrections and possible colored signals even within these models of neutral naturalness, some of which might appear at the LHC or future colliders.}
\begin{document} 
\maketitle
\flushbottom

\section{Introduction}
As the LHC has continued to run at 13 TeV, the absence of any clear sign of new physics has strained traditional models of naturalness. Models such as Supersymmetry \cite{Martin:1997ns}, Little Higgs\cite{ArkaniHamed:2002qy}, or extra dimensions/compositeness \cite{Randall:1999ee} all predict colored top-partners at the TeV scale. While it remains a possibility that such states exist but remain out of kinematical reach, the motivation to consider models of ``neutral naturalness'' grows.

The most prominent such scenario is that of the Twin Higgs \cite{Chacko:2005pe}. In it, the Higgs is a pseudo-Goldstone boson of an  (approximate) SU(4) symmetry in the theory.
Specifically, one considers a Lagrangian 
\bea\label{LTH}
\mathcal{L}&=\lambda_H\left(|H_A|^2+|H_B|^2-f^2\right)^2+\hat{y}_e\left(H_A^\dagger \ell e_R^c+ H_B^\dagger L E_R^c\right)\\
&+\hat{y}_u\left(H_A q u_R^c
+ H_B Q U_R^c\right)+ \hat{y}_d \left(H_A^\dagger q d_R^c +H_B^\dagger Q D_R^c\right).
\eea
After breaking of the $SU(4)$, the light degrees of freedom are identified as the Higgs bosons, and we are not quadratically sensitive to the cutoff scale.
The theory contains top-partners, but because they are uncolored, they naturally evade detection at the LHC. 

Of course, this success comes at a cost. The theory requires the complete doubling of the degrees of freedom of the standard model. Such an extension is not at all unprecedented, with SUSY also doubling the spectrum into superpartners. Like SUSY, the Twin symmetry must be broken at some level, but {\em unlike} SUSY, the doubling does not come with the same questions of flavor, baryon and lepton number conservation.

Among other reasons that one must break the twin parity is the cosmological constraints on the theory. The presence of twin neutrinos and twin photons in the universe is strongly constrained, with Planck recently limiting the presence of additional relativistic species  $\Delta N_{eff} < 0.30$ at 95\% confidence level\cite{Aghanim:2018eyx}. 

 One might attempt to solve this problem simply by fiating away the unwanted degrees of freedom. This, however, fails because these light degrees of freedom are simply the repository of the entropy of the twin sector. If the these degrees of freedom are removed, the entropy will merely reside in heavier particles, such as twin pions. One possibility is to dilute the entropy of the twin sector by a late reheating, as in e.g., \cite{Chacko:2016hvu}. Absent that, one must find a way for that entropy to find its was back to the SM.
 
 This is a challenge because of the limited interaction of the Twin sector to the SM. Indeed, the success of the Twin Higgs scenario arises largely because the coupling to the SM is so weak (arising through the Higgs portal). The presence of an additional portal to the SM may be the needed ingredient to make these models cosmologically safe. Already, the kinetic mixing portal has been explored \cite{Hochberg:2018vdo}. Here, we explore the neutrino portal, as suggested in \cite{Schmaltz:2017oov}.
 
 The core idea is to consider a singlet fermion, deemed a ``portalino,'' which does not transform under the $Z_2$ that exchanges the different sectors. As argued in \cite{Schmaltz:2017oov}, such an object marries a single linear combination of the SM and Twin neutrinos, lifting that combination and leaving a massless combination that we identify with the SM neutrino. Such an object is a general category of ``singletons'' \cite{Bishara:2018sgl}, which have a rich phenomenology in Twin models.

However, this naturally yields a mixing angle $O(v/f)$ between the SM and twin neutrino, which is in general too large. Thus, some modifications to the theory must be included. As we shall see,  models with limited additional matter naturally address this. Moreover, their existence  yields a set of important experimental signals of such a scenario.

The layout of this paper is as follows: we shall begin by showing how to approach this "fiat" solution where light fields are lifted with spontaneous $Z_2$ breaking. This will leave (at a minimum) a single light neutrino species. We shall discuss how the portalino in marrying the neutrinos begins to address these problems, but still remains highly constrained. We then will show how the extension of the theory to a Twin-2HDM model (specifically Type -II or Type-III)  naturally allows a simple resolution of all cosmological problems in sec.~\ref{sec:weak}. As an alternative approach, the cosmological problem can be solved straightforwardly if the $\ztwo$ asymmetry lifts light fermions while breaking Twin color. We will propose a so-called strong solution along this direction in sec.~\ref{sec:strong}. Finally, we discuss the experimental implications and conclude.

\section{A First Step to Lifting the Light Degrees of Freedom from Spontaneous Twin $U(1)$ Breaking}\label{sec:u1break}
To begin, we should discuss how we might arise at a simple theory that ameliorates the light degree of freedom problems. Allowing a transfer of entropy to the SM is not enough if we still have massless degrees of freedom. We shall see this is easily realized with spontaneous $Z_2$ violation in the theory.

We begin by adding a field $\phi$ to the SM, with $Y=1$, and its twin partner. We can then include a term in the theory

\begin{equation}\label{Vphi}
V(\phi,\Phi) = -m^2_\phi (|\phi|^2 + |\Phi|^2) + \lambda_{\phi}(|\phi|^4 +|\Phi|^4) + \kappa_{\phi} |\phi|^2 |\Phi|^2\;
\end{equation}
Such a potential respects the $Z_2$, and thus critically will not even radiatively spoil the $\ztwo$ symmetry of the quadratic mass terms. 
We can also consider  tree level quartic terms $ -\alpha_\phi\left(|\phi H_A|^2 + |\Phi H_B|^2\right)$, which do break the global $SU(4)$. To maintain an approximate global symmetry, we are limited to $0< \alpha_\phi \ll 1$. Because this is {\em only} a $\ztwo$ and not a continuous symmetry, when $\Phi$ acquires a vev  $\langle\Phi\rangle = v_\phi$ and breaks the \ztwo, the field $\phi$ retains a (positive) TeV-scale mass.

Once this occurs, the photon in the twin sector is massive, and at least part of the d.o.f. problem is solved. Although this solution is fairly trivial, it has not, to our knowledge, been actually written down in the literature at this point.

With the presence of a $\Phi$ vev, we can begin to contemplate how that affects other degrees of freedom. In particular, we can lift two $SU(2)$ doublets with a term $\cal L$ $\supset Y_\phi \Phi L_i L_j \epsilon^{ij}$, while the third remains in the theory, protected by the Witten anomaly. 

We still have a massless neutrino and two massless right-handed leptons in the theory at this point. To lift these, we can now introduce the portalino field $N$, which will marry $Y_E\Phi E^c N$. With two portalino fields one can envision lifting the twin $\mu$ and $e$ fields, leaving the massive twin $\tau$ and only a single massless neutrino. This might seem tantalizingly close to a solution to the problem, but as we have already alluded to, even if all these degrees of freedom were lifted, the theory would still have pions. 

To deal with these, one must move the entropy into the SM. If there exists sizable kinetic mixing between visible photon and twin photon, then this problem many be solved in the SIMP framework~\cite{Hochberg:2018vdo}. Absent such a mixing, something else must solve the problem of Twin Entropy. Twin pions are naturally unstable with twin-hypercharge broken. However, their lifetimes are  quite long. Indeed, with all but the twin neutrino lighter than the pions, the twin pions will decay via
the $\phi$ mediated channel $\Pi \rightarrow e_i^+ e_j^- \bar{\nu_k}\nu_s$. This process is suppressed by both large $\phi$ mass and the phase space volume, 
and the decay width turns out to be exceedingly small (less than $10^{-40}$ GeV for the typical model parameter values in this work).    


If the twin pion entropy is to be depleted via decay, this points to a scenario where at least one massive left-handed twin lepton is lighter than twin pions, so that pions are short-lived. However, as we have already said, if there are still massless degrees of freedom in the theory, the problem of twin entropy remains.

\subsection{An aside on reheating}
In this setup, the decay of $N_{1, 2}$ is preferentially to SM final states. Thus, the presence of cosmologically produced $N$ states reheats the SM sector and effectively dilutes the energy density of the twin sector. This motivates us to consider an alternative scenario which also avoids conflicts with constraints on $\neff$. To be specific, if there were a time in the early universe that $N_{1, 2}$ came to dominate the energy density and soon asymmetrically decayed in such a way that the contribution of the massless twin neutrino energy density is suppressed by one order of magnitude, the current cosmological bounds on dark radiation will be satisfied~\cite{Chacko:2016hvu}. In such a reheating scenario, most of $N_a$'s decays should take place after A-B sectors thermally decouple and before BBN. However, this leads a very stringent constraint on model parameters, namely, $9.4\times 10^{-4}<Y_E<2.7\times 10^{-3}$ if one fixes $v_\phi = 10\; \tev$ and $Y_\phi = 10^{-2}$. We will present the details in the appendix~\ref{appx}. Thus, one can see a complete, albeit tuned model, which motivates a study of a less tuned model. We leave this for future work.

\section{A Portalino to the Twin Sector}\label{sec:portalino}
Let us now take the previous section and expand and generalize it. Let us consider the Twin Higgs scenario, augmented with the additional fields $\phi$, and further augmented by three portalino fields $N$. By portalino, we assume here that they will marry any gauge singlet fermions that exist in the low energy theory, where a gauge singlet dimension$-5/2$ operator exists in the UV, including the twin neutrinos $L H N$ and the twin right handed fermions $\Phi E^c N$. 

Taking in all these concerns, the Yukawa interactions which involve $N$, $\phi$ and $\Phi$ are
\bea\label{LTP}
\mathcal{L}\supset\left(H_A\ell+H_B L\right)\hat{Y}_{\nu} N+\left(\phi^\dagger e_R^c+\Phi^\dagger E_R^c\right) \hat{Y}_{E} N+\phi \ell\hat{Y}_\phi \ell+\Phi L\hat{Y}_\phi L.
\eea
The dimensionless coupling parameters $\hat{Y}_{\nu}$, $\hat{Y}_{E}$ and $\hat{Y}_\phi$ are $3\times 3$ matrices. The most salient feature for our purposes is the lifting the light twin degrees of freedom. In this sense, although the flavor structures of these Yukawa matrices could be arbitrarily complicated, typically only their eigenvalues matter. In order to generate minimal sources of lepton flavor violation processes at low energy regime, we choose $\hat{Y}_\nu$ and $\hat{Y}_E$ being diagonal in the interaction basis, and assume that such an approximate structure can be suitably realized by a UV flavor symmetry, see e.g., \cite{Egana-Ugrinovic:2018znw}. Meanwhile, $\hat{Y}_\phi$ can be in general expressed in terms of Levi-Civita tensor since it has to be antisymmetric which has at most three independent components, i.e.
\bea 
\hat{Y}_{\nu}=\text{diag}(Y_{\nu_1}, Y_{\nu_2},Y_{\nu_3})\;,\quad \hat{Y}_{E}=\text{diag}(Y_{E_1}, Y_{E_2},Y_{E_3})\;,\quad \hat{Y}_\phi=\frac{1}{2}Y_{\phi_k}\epsilon^{kij}\;,
\eea
After $\ztwo$ breaking, the VEV of $\Phi$ will contribute the masses of the twin leptons, and their mass spectrum structure become different from their SM partners. In this subsection, we will resolve the twin lepton mass spectrum under the assumption $Y_{\phi_1}=Y_{\phi_2}=0$, $Y_{\phi_3}\neq 0$, the more general case will be discussed in the appendix~\ref{appx}. 
As we will see later, in the presence of a portalino there will be one massless eigenstate which is consisted of the linear combination of $L_3$ and SM left-handed neutrino with mixing angle $\alpha \sim(v_h/f)$. As we shall see shortly, experiment constrains this to be small, while naturalness suggests this should be large. This tension determines what parameter space exists, although in regions with even moderate naturalness ($v/f>0.1$) the allowed parameter space is small.

In attempting to determine what is allowed, we focus on concrete regions of parameter space. The "tau neutrino mixing" parameter $|U_{\tau n}|$ has the least stringent constraint, thus we consider the case where the first two generations are lifted with $\Phi$. In the first two generations the Lagrangian e.q. (\ref{LTP}) takes the following form,
\bea\label{Ya}
\mathcal{L}_{Y_a}&=Y_{E_{a}} v_\phi \left(E_{R_a}^\dagger +U_{e_a n}\nu_a\right) N_a+Y_{\nu_{a}}f\hat{\nu}_a N_a+Y_{\phi_3} v_{\phi} \hat{\nu}_a\epsilon^{ab} E_{L_b}\\
\eea
where $U_{e_a n}=Y_{\nu_a}v_h/(Y_{E_a} v_\phi) $.  We take $v_\phi\gg f > v_h$, and thus the neutrino mixing angle $|U_{e_a n}|$ can be sufficient small to evade the experimental constraints. In the mass basis $\mathcal{L}_{Y_a}$ is equivalent to two Dirac mass terms say, $M_a^+\Psi_{E_{R_a}}\Psi_{N_a}$ and $M_a^-\Psi_{\hat{\nu}_a}\Psi_{E_{L_a}}$
where $M_a^\pm$ are the mass eigenvalues,
\bea
\left(M_a^\pm\right)^2=\frac{\mu_a^2}{2}\left(1\pm\sqrt{1-\frac{4Y_{E_a}^2 Y_{\phi_3}^2 v_\phi^4 }{\mu_a^4}}\right)\;,\quad \mu_a^2=(Y_{\phi_3}^2+Y_{E_a}^2)v_\phi^2+Y_{\nu_a}^2 f^2\;,
\eea
note that the Einstein summation rule is not applied in this expression. The corresponding mass basis $(\Psi_{\hat{\nu}_a}, \Psi_{E_{R_a}})$ and $(\Psi_{N_a}, \Psi_{E_{L_a}})$ are rotated from the Yukawa interaction basis
through the angular parameters $\alpha_a$ and $\beta_a$ respectively,
\bea
\bmtx \Psi_{\hat{\nu}_a}\\ \Psi_{E_{R_a}}\emtx = \bmtx c_{\alpha_a} & -s_{\alpha_a} \\ s_{\alpha_a}& c_{\alpha_a}\emtx \bmtx \hat{\nu}_b \\\widetilde{E}_{R_a}^\dagger  \emtx \;,\;
\bmtx \Psi_{N_a}\\ \Psi_{E_{L_a}} \emtx = \bmtx c_{\beta_a} & -s_{\beta_a} \\ s_{\beta_a} & c_{\beta_a} \emtx \bmtx N_a \\\epsilon^{ab}E_{L_b}\emtx 
\eea
where $\widetilde{E}_{R_a}^\dagger=E_{R_a}^\dagger +U_{e_a n}\nu_a$ and the rotation angles can be determined from the relation: $\tan(\alpha_a\pm\beta_a)=Y_{\nu_a}f/(Y_{E_{a}}v_\phi\pm Y_\phi v_\phi)$.
At the same time, we also obtain two massless modes
\bea
\psi_{\nu_a}=\nu_a -U_{e_a n}E_{R_a}^\dagger \approx \nu_a.
\eea
In the third generation, $L_3$ mass is given by the VEV of $H_B$ while the Dirac mass $E_{R_3}$ and $N_3$ is sourced from $v_\phi$ which could be much heavier. Since there would be a mass hierarchy between those fermions, it is most convenient to establish the effective Lagrangian for the third generation twin leptons.
To be specific, if $Y_{E_3}>0.1$, $E_{R_3}$ and $N_3$ get a Dirac mass $M_{N_3}\simeq Y_{E_3} v_\phi $ and can be   integrated out from their e.o.m.. As a result, one arrives the third twin lepton mass term $M_3\Psi_{\hat{\nu}_3} E_{L_3}$ where, $M_3=m_\tau Y_{\nu_3}f/[M_{N_3}\sin(\alpha_3)]$, $\Psi_{\hat{\nu}_3}=\cos(\alpha_3)\hat{\nu}_3+\sin(\alpha_3) \nu_3$ and $\alpha_3=\arctan(v_h/f)$. $\Psi_{\hat{\nu}_3}$ is supposed to be the lightest twin particle and typically we will need
$M_3^2 < (M^\pm_a)^2$ to guarantee this. We could assume the value of $M_3$ is around the muon mass, so that $\Psi_{\hat{\nu}_3}$ will decay into SM leptons very fast by the virtue of sizable mixing angle provided by this portal right-handed neutrino model, 
\begin{equation}
\Gamma(\text{Twin}\rightarrow\text{SM})\simeq \frac{\sin^2(2\alpha_3)}{ 9\times 10^{-6}\text{sec}}
\end{equation}
The cosmology problem will be solved if the twin particle are depleted before BBN. Since $E_{L_3}$ component of $L_3$ is massive, the leftover degree of freedom 
\bea
\quad \psi_{\nu_3}=\cos(\alpha_3) \nu_3 -\sin(\alpha_3)\hat{\nu}_3
\eea
which is orthogonal to $\Psi_{\hat{\nu}_3}$ becomes the third massless neutrino. From the above discussion, we see that the mixing mixing strength $|U_{\tau n}|^2=\sin^2(\alpha_3)$  between the $\nu_\tau$ and the twin neutrino (the sterile neutrino) is sizable when the parameters are natural, i.e. $\tan\alpha_3\approx v_h/f\gtrsim 0.1$. In a more general context that $Y_{\phi_{1, 2}}\neq 0$, the mixing angle will be enhanced as $\tan(\alpha_3)=v_{h}\sqrt{Y_{\phi_i}^2}/(f Y_{\phi_3})$ which makes the situation worse. 

With this understanding of the mixing angle and naturalness, we can confront the basic constraints. We would point the reader to ~\cite{Orloff:2002de,Atre:2009rg,Helo:2011yg, Bertoni:2014mva,deGouvea:2015euy,Batell:2017cmf,Schmaltz:2017oov} for a broad discussion. For our particular region of parameter space, the CHARM experiment is most constraining, setting an upper bound on the mixing angle with the sterile neutrino as a function of mass. This limit monotonically gets stronger as the sterile state gets heavier in our region of interest. For example, if $|U_{\tau n}| \leq 0.08$ we can have states $\overline{M}_3= 50$ MeV, and if $|U_{\tau n}| \leq 0.32$ the sterile state can be no heavier than $20$ MeV;  on the other hand the portalino masses smaller than 10 MeV are ruled out by the CMB bound on extra relativistic degrees of freedom $\neff$. As a consequence, there is limited space for the model to exist. Generously taking $f/v_\phi =0.1$, and $|U_{\tau n}|= 0.17$ we then have $10^{-2}<Y_{\nu_3}/Y_{E_3}<3.5\times 10^{-2}$, which implies a small window in the parameter space for the model.

\subsection{Low Energy Constraints}
Before moving on to models that can address this, it is worth discussing the low energy constraints that apply generally to these sorts of scenarios. 
In order to provide sizable twin lepton masses, the Yukawa coupling of the charged scalar $Y_{\phi_i}$ cannot be too small.
In general, any flavor non-diagonal coupling will induce various lepton flavor violation (LFV) decay processes and at the same time yield contribute to the anomalous magnetic moments of  leptons. Coincidentally $\phi$ involves the same Yukawa interaction as the singly charged scalar in Zee-Babu model\cite{Babu:1988ki, Babu:2002uu}. This means we can  adopt the  experimental limits from earlier studies. 

Specifically, we have:
\begin{enumerate}[label=(\roman*)]
\item{ $\ell_a \rightarrow \ell_b \nu\bar{\nu}$: The Fermi muon decay constant will be affected when this process is considered\cite{Nebot:2007bc, Schmidt:2014zoa}, 
\bea
\left(\frac{G_{\mu}^{eff}}{G_{\mu}^{\text{SM}}}\right)^2\approx 1+ \frac{\sqrt{2}}{G_F M_\phi^2}|Y_\phi^{e\mu}|^2
\eea
where we show only the leading terms in the couplings $Y_\phi^{e\mu}$, which emerge from the interference of the SM diagram with the ones mediated by $\phi$. Since the charged scalar only contributes to lepton decays but not hadronic decays, the universality of the couplings in hadronic and leptonic decays can be tested by assuming the unitarity of the  Cabibbo-Kobayashi-Maskawa (CKM) matrix in the framework of SM, i.e, $G_{\mu}^{eff} V_{ij}^{\text{exp}}=G_{\mu}^{\text{SM}} V_{ij}$, where $V_{ij}$ are the truly elements which gives,
\bea
\sum_{q= u, d, s}|V_{uq}^{\text{exp}}|^2=1- \frac{\sqrt{2}}{G_F M_\phi^2}|Y_\phi^{e\mu}|^2= 0.9999\pm 0.0006
\eea
In fact, such relation provides the strongest bound on our model parameter which translates as $|Y_\phi^{e\mu}|^2 < 0.014 (M_\phi/\tev)^2$.}\\
\item{ $\mu\rightarrow e\gamma$ and $\mu N - e N$ conversion in nuclei. These two processes are highly related and the former has a simpler explicit expression of the branching ratios, BR($\ell_i\rightarrow\ell_j\gamma$)=$R_i^{j\gamma}\times \text{BR}(\ell_i\rightarrow\ell_j \nu\bar{\nu})$, where $i\neq j$ where,
\bea
R_i^{j\gamma}=\frac{\alpha_{em}}{48\pi}\left |\frac{(\hat{Y}_\phi\hat{Y}_\phi)_{ij}}{G_F M_\phi^2}\right |^2
\eea
The constraint from $\mu\rightarrow e\gamma$ gives rise to $|Y_\phi^{e\tau}Y_\phi^{\tau\mu}|^2 < 1.1\times 10^{-6} (M_\phi/\tev)^4$. $\mu-e$ conversion is induced from the effective $\mu e\gamma$ vertex and the relevant expressions can be found in~\cite{Lindner:2016bgg, Schmidt:2014zoa}. Current experimental limts come from Ti and are slightly worse than present $\mu\rightarrow e\gamma$ constraints. One should note that projected limit is supposed to be tighter when the nucleus of Aluminum are used in the near future.}
\item{ For diagonal transition we will resort the constraint form muon and electron anomalous magnetic moments: $a_\mu = (g-2)/2$ with,
\bea
\Delta a_\mu \equiv a_\mu^{\text{exp}}-a_\mu^{\text{SM}}=- \frac{|(\hat{Y}_\phi\hat{Y}_\phi)_{\mu\mu}|^2m_\mu^2}{24\pi^2 M_\phi^2}
\eea 
Since the experimental measurement of $a_\mu$ is slightly larger than the SM prediction~\cite{Schmidt:2014zoa}, we will only use this information to derive constraints on our model parameter space. By demanding the contribution with in the error bar $\Delta a_\mu^{err}\sim 10^{-11}$, we simply get $|Y_\phi|<0.5 M_\phi/\tev$.}
\end{enumerate}
\subsection{A weak solution}\label{sec:weak}
The $U(1)_Y$ scalar model is a simple example of neutrino portalino mechanism with direct energy from the twin to the SM sector. However, as we have already realized, this simplest model is tightly constrained. There is no mystery as to the origin of this - the mixing angle for the massless ``neutrino'' state is a ratio of Dirac masses, and Twin symmetry requires this angle to then be $\alpha \sim v/f$. 

But Twin symmetry is broken, and in particular is broken here by the Twin $U(1)$ $\Phi$ vev. As a consequence, one needn't necessarily have the ratio of masses be $v/f$.  

\begin{figure}[t]
\centering
\includegraphics[width=0.6\textwidth]{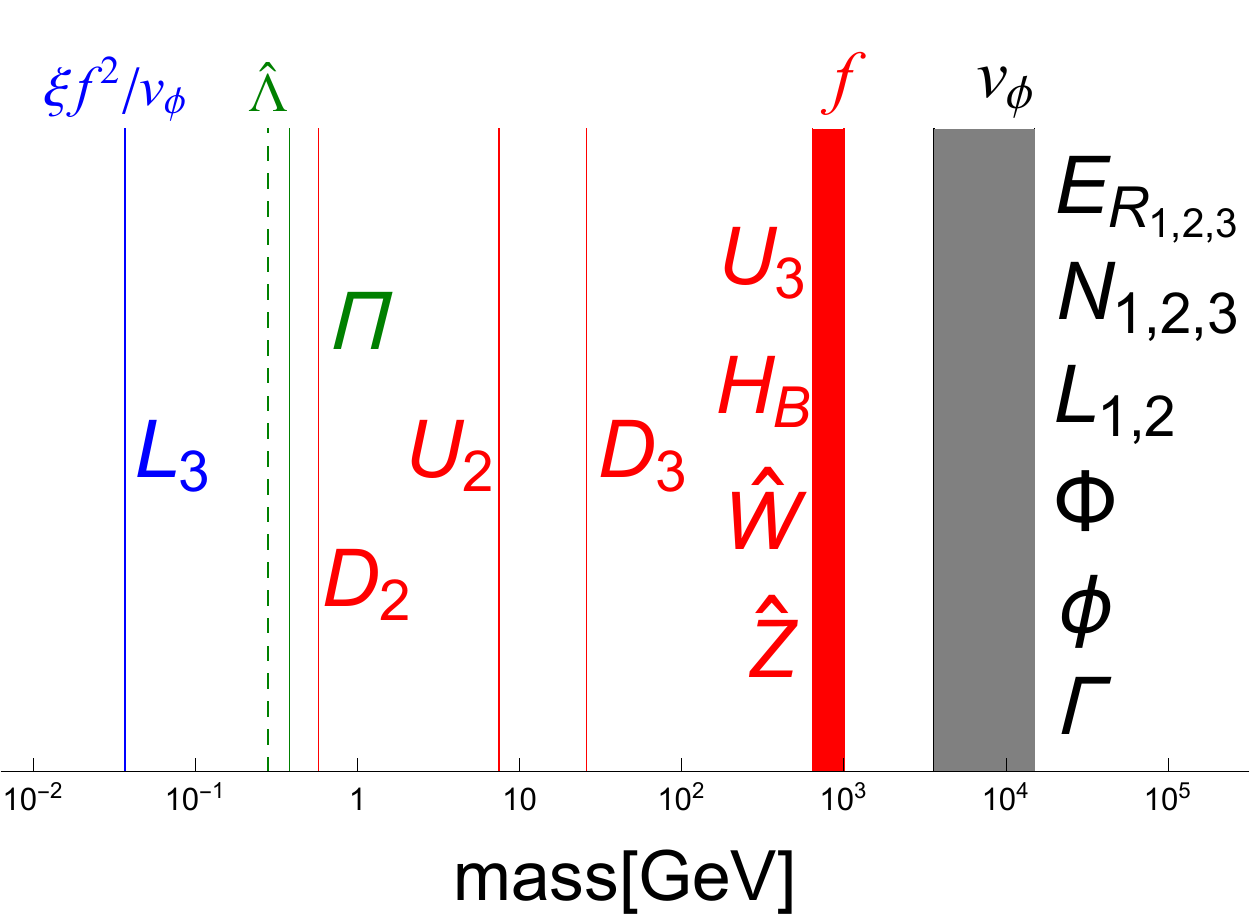}
\caption{A schematic of the spectrum of the theory with the field content in table \ref{tb:weaksolncharge}.}
\label{fig:weakspectrum}
\end{figure}

For instance, consider the following dim-5 operators in the low scale theory,
\bea\label{t2eft}
\mathcal{L}_\eff =\frac{1}{M}\left[\zeta_1 \left(\phi^\dagger H_A \ell e_R^c+\Phi^\dagger H_B L E_R^c\right)-\zeta_2 \left(\phi H_A^\dagger\ell +\Phi H_B^\dagger L\right) N\right]
\eea
where $M$ is a mass scale and $\zeta_i$ are dimensionless coefficients. Replacing $\Phi$ by its VEV and inserting $\mathcal{L}_\eff$ into the third family leptonic interactions we then arrive at
\bea\label{lowYuk}
\mathcal{L}_3 = \left(y_{e_3} H_B^\dagger +C_{\phi_1} \epsilon H_B \right)L_3 E_{R_3}^c+ \left[Y_{\nu_3}H_A \ell_3 + \left(Y_{\nu_3}\epsilon H_B - C_{\phi_2}H_B^\dagger \right)L_3\right]N_3,
\eea
where $C_{\phi_i} =\zeta_i v_\phi/M$. We immediately find that the SM tau neutrino mixing angle is then modified as $U_{\tau n}= Y_{\nu_3} \tan\alpha_3/C_{\phi_1}$ in the limit of  $C_{\phi_2}\gg Y_{\nu_3}$.
At high energy scale, those dim-5 operators can be induced from additional $SU(2)_L$ doublet scalar fields $ \Sigma$ in the Twin sector, i.e. we will have extended the Higgs content to two-Higgs doublet models (2HDM) in each sector. We  summarize the additional ingredients for the weak solution in the  table \ref{tb:weaksolncharge}. A schematic of the resulting spectrum we show in Figure \ref{fig:weakspectrum}.

\begin{table} 
	\begin{center}
  \begin{tabular}[t]{ |c | c | c | c|}
  \hline
  Names & $\Sigma$ & $\Phi$ &$N_i$ \\ 
   \hline
 $[SU(2)_L\times U(1)_Y]_{B}$ & $\left(2, +\frac12\right)$ & $(1, 1)$ & $(1, 0)$ \\
   \hline
   $Q_{L}$ & $0$& $-2$&$-1$\\
   \hline
  \end{tabular}
\end{center}
\caption{Additional twin fields shown for model described in section \ref{sec:weak}. (Related SM fields not shown.)}
\label{tb:weaksolncharge}
\end{table}

2HDMs are cataloged as various types depending on the Yukawa coupling terms. Taking the general form of the leptonic interaction the additional Lagrangian associated with the new fields is given by, 
\bea
\mathcal{L}_\Sigma &\supset -M_{\Sigma}^2|\Sigma|^2 -\mu_\sigma^2 H_B^\dagger\Sigma + m_\sigma \Phi^\dagger H_B\Sigma  +\left(\tilde{y}_\tau H_B^\dagger+ g_\tau \Sigma^\dagger \right)L_\tau \tau_R^c +\ztwo ^{\text{twin}}\\
&+\left(\tilde{Y}_{\nu_3} H_B+ g_{\nu_3}\Sigma\right) L_\tau N_3 
\eea
here and in what follows $\mathbb{Z}_2^{\text{tw}}$ refers to the twin counterparts. Before considering the $\ztwo$ symmetry breaking, we note that the mass-squared matrix of $(H_A, \sigma)$ and $(H_B, \Sigma)$ can be simultaneously diagonalized and the parameter $\theta$ is defined as the rotation angle that performs that diagonalization,
\begin{equation}
\tan(2\theta) =\frac{2 \mu_\sigma^2}{2\lambda_h f^2+M_\Sigma^2}
\end{equation}
The mass eigenvalues depends on $\theta$, to maintain the relation $\langle H_B\rangle\approx f$ we require $|\theta| \ll 1$ in our model.  For $M_\Sigma > M_\phi$, we can obtain $\mathcal{L}_\eff$ by integrating out $\Sigma$ and yield $\zeta_{1, 2}=g_{\tau,\nu_3}m_\sigma M/M_{\Sigma}^2$. 
Even if $M_\Sigma \lesssim M_\phi$, this solution is valid as well.
Since in the ground state of $\Phi$, the $m_\sigma$ trilinear interaction reduces to an off-diagonal mass term $m_\sigma v_\phi H_B\Sigma$ which can be got rid of via rotation angle $\gamma$, where $\tan(2\gamma) = m_\sigma v_\phi\tan(2\theta) /\mu_\sigma^2$. The resulting VEV of $\Sigma$ takes the form of $\langle \Sigma\rangle \simeq (\gamma, \theta)^{T}f $, therefore yields extra mass terms for $E_{L_3}$ and $\hat{\nu}_3$ so that $U_{\tau n}$ is suppressed. Since this quadratic term does not appear in the visible sector, $m_\sigma v_\phi$ will contribute to mass of $H_B$. To aviode spoiling twin Higgs mechanism, it is crucial to have $|\gamma| < 1$. The tau-neutrino mixing angle takes different expressions which is determined by types of 2HDM. For later convenience, let's define a suppression factor $\kappa\equiv U_{\tau n}/\tan\alpha_3$, i.e., a measure of how small the mixing angle is compared to the naive value of sec \ref{sec:portalino}. 
In type I or II 2HDM only one of the Higgs doublet couples to the leptons in the interaction basis. We find that $\kappa$ is proportional to $\cot\theta\cot\gamma$ in type a I model and is proportional to $\tan\theta\cot\gamma$ in type a II model at leading order. 
As to the type III 2HDM, both of the two Higgs bosons directly couple to fermions~\cite{Branco:2011iw}, $\kappa = \tilde{Y}_{\nu_3}\cot\gamma/ g_{\nu_3}$ even though we set $\tan\theta =0$.
For $\tilde{Y}_{\nu_3}$ and $g_{\nu_3}$ are treated as independent model parameters, small $\tilde{Y}_{\nu_3}$ will lead to $|\kappa| < 1$. 
In this sense, the type II or type III case is more promising to provide small $\kappa$. For the type II 2HDM is the most studied one, since the couplings of the MSSM are a subset of the couplings of the type II 2HDM. We will go through more details about this set up and see how the cosmological problem is solved.

As discussed in previous section, the first two generation twin leptons can be decoupled from the third generation since they may obtain large masses from $\langle\Phi\rangle$.  In the case of $M_{N_3}=Y_{E_3}v_\phi>f$, $N_3$ and $E_{R_3}$ are decoupled as well. Consequently, the low scale effective theory 
provides twin lepton mass terms,
\bea\label{t2mass}
\mathcal{L}_{\eff}&= M_{\Psi_3}\left(\sqrt{1-\kappa^2}\hat{\nu}_3-\kappa E_{L_3}\right)\left(\sqrt{1-\kappa^2}E_{L_3}+\kappa\hat{\nu}_3-\kappa\tan(\alpha_3)\nu_3 \right)
\eea
where $M_{\Psi_3}\simeq m_\tau g_{\nu_{3}}\gamma f /[M_{N_3}\kappa \tan(\alpha_3)]$
and $\kappa \simeq \theta/\sqrt{\theta^2+\gamma^2}\simeq \mu_\sigma^2/(m_\sigma v_\phi)$.
There are two orthogonal linear combinations of $\hat{\nu}_3$, $E_{L_3}$ and $\nu_3$ that share a Dirac mass in e.q.~(\ref{t2mass}); for $\kappa^2 \ll 1$ we denote the first combination as $\Psi_{\hat{\nu}_3}$ and the second one as $\Psi_{E_{L_3}}$. In this sense, the lightest massive twin particles are $\hat{\nu}_3$ and $E_{L_3}$ for their masses are suppressed by $M_{N_3}$ through the seesaw mechanism and therefore become the final states of twin pion decays. In addition to that, there is a massless mode $\psi_{\nu_3}\simeq\nu_3+U_{\tau n}E_{L_3}$ which would be identified as physical neutrino. We assume $|\kappa|$ is not too small but less than unity so that it can provide a sufficient decay rate to convey hidden entropy to visible sector via the admixture between twin leptons and SM neutrinos in the mass eigenstates. Such mass spectrum for light twin leptons can satisfy the constraints from both cosmological observation (e.g. BBN) and current experimental results (e.g. CHARM) in the naturalness regime. For example, if taking the following values,
$g_{\nu_3}=4.0\times10^{-2}$, $\tan\theta=0.04$, $\tan\gamma=0.3$, $f=1\;\tev$ and $M_{N_3}=8.0\;\tev$
we then get $M_{\Psi_3}=117\;\text{MeV}$ and $ |U_{\tau n}|=2.28\times 10^{-2}$. 
Those twin leptons will further decay into SM states via the 4-Fermion operator $\mathcal{O} \simeq U_{\tau n}G_F (\Psi_{E_{L_3}}^\dagger\bar{\sigma}^\mu\psi_{\nu_3})(\psi_{\nu_i}^\dagger \bar{\sigma}_\mu\psi_{\nu_i})/\sqrt{2}$ which gives rise to the dominating partial decay width $\Gamma (\hat{\nu}_3, E_{L_3}\rightarrow 3 \nu)=1/(2.57\times 10^{-3}\text{sec})$.

In summary, while the details can get buried in mass matrix diagonalization, it is not challenging to realize a model where the naive tensions of section \ref{sec:portalino} are not present. In particular, the presence of additional Higgs fields can produce mass terms after $\ztwo$ breaking that differ between twin and SM sectors. As a consequence, there can be small, but non-trivial mixing between the SM neutrino and Twin fermions. All Twin states are massive, and the decays ultimately transfer the entropy of the Twin sector back to the SM.

\subsection{Constraints from Neutrino Oscillation}
Two Higgs Doublet Models are  some of the most popular extensions of SM. As a result, experimental constraints on models including two Higgs doublets  - in almost any form - can be find in many articles e.g.\cite{Branco:2011iw}. With the presence of new scalar fields $\phi$, the visible part of our model is identical to the Zee model\cite{Zee:1985rj, Zee:1985id} which has been studied in several literatures \cite{Balaji:2001ex, Herrero-Garcia:2017xdu}. In Zee model the coexistence of the two Higgs doublets in the $m_\sigma$ term is required to generate the mass terms for neutrinos and explicitly violates the lepton number. We can see the origin of neutrino mass terms from loop induce Weinberg operator in the diagram below.\\
\begin{figure}[t]
\centering
\includegraphics[width=0.4\textwidth, height=0.5\textwidth]{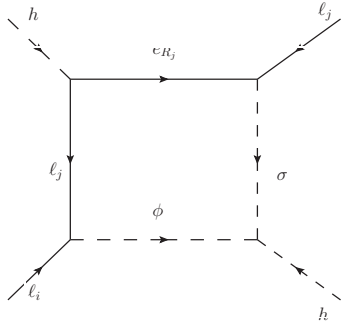}
\caption{Box diagram that generates Weinberg operator.}
\end{figure}
\\
In terms of model parameters, the diagram is given by,
\bea
\mathcal{O}_{wein}&= \frac{m_\sigma s_{2\theta}}{64\pi^2(M_\phi^2-M_\Sigma^2)}\ln\left(\frac{M_\phi^2}{M_\Sigma^2}\right) \left(\hat{g}_e^2\hat{Y}_\phi\right)^{ij}\left(\ell_i h)(h\ell_j\right)
\eea
The simplest case comes from the assumption that $\phi$ only couples to the first two generation leptons and the neutrino mass is the mixing term between $\nu_e$ and $\nu_\mu$. Correspondingly, the main contribution to the flavor off-diagonal neutrino mass is estimated as,
\bea
\mathcal{O}_{wein}= 0.1\text{eV}\left(\frac{m_\sigma}{100\gev}\right)\left(\frac{Y_\phi}{3\times 10^{-3}}\right)\left(\frac{0.1}{\tan\theta}\right)\left(\frac{1\tev}{M_\phi}\right)^2\nu_e\nu_\mu
\eea
Although the original version of the Zee model predicted neutrino mass is disfavored by neutrino oscillation data~\cite{Jarlskog:1998uf,Koide:2001xy, Balaji:2001ex, Oliver:2001eg}, the generalized version (embedded in the type-III 2HDM) is still a promising one. In our case, due to the presence of right-handed portalinos and nontrivial admixture with SM singlets from twin sector, the neutrino mass matrix structure would be more flexable and the details will left for future work. Currently, we will regard this l-loop neutrino oscillation term as a constraint from lepton flavor violation (LFV).  
\section{A strong solution}\label{sec:strong}
It is possible that the twin $\mathbb{Z}_2$ SSB occurs in the strong sector, i.e. twin $SU(3)_c$ is broken at the ground states of QCD charged scalars. In this section, we will pursue such a model.

As we shall see, the cosmology of this setup is markedly different from the weak one. Rather than eliminating the light fields, we shall see they can naturally decouple early. 
If twin color is broken completely, i.e., $SU(3)_B \times SU(2)_B \times U(1)_B \rightarrow_{\vev{X,\tilde X}} SU(2)_B \times U(1)_B \rightarrow_{\vev{H_B}} U(1)$, then the quarks and leptons of the sector can all be lifted, either through mutual couplings or couplings to the portalino. 

The fermion fields can naturally be heavier than the bottom quark of the SM, but maintain equilibrium through the Higgs portal. The final entropy after decoupling of the twin sector is much less than that of the SM, and thus the light degrees of freedom are not a problem, even without a late reheating.

In order to realize this scenario, there must be at least two colored fields $\tilde{X}$ and $X$ (with their twin partners) in order to full break color. This both allows us to fully lift the twin fermions. Moreover, if the theory had confining gluons, the massive glueballs would overclose the universe.) For our realization, we will need a partial misalignment of vevs of the fields $X, \tilde X$, although in more complicated models multiple fields to serve this role as well. The field content beyond a standard Twin Higgs + Portalino scenario is shown in table \ref{tb:strongsolncharge}.

We shall require sizable quark-quark Yukawa terms, which will imply that twin baryon number is broken. For the obvious requirements of the stability of the SM proton, this must occur along with the $Z_2$ breaking as well. This will be achieved with a pair of gauge neutral but $[U(1)_{\text{bn}}]_{A, B}$ charged scalar fields $s$ and $S$ to align the vacuum orientations in such a way that $U(1)_{\text{bn}, B}$ is spontaneously broken while $U(1)_{\text{bn}, A}$ is not. In the same spirit of $U(1)_Y$ model, we replace the $[U(1)_Y]_{A, B}$ charged scalar fields in e.q.~(\ref{Vphi}) with $s$ and $S$,
\bea
V\supset\lambda_s \left(|s|^2+|S|^2-v_s^2\right)^2+\eta_s|s|^2|S|^2
\eea
where $\lambda_s, \eta_s > 0$ and $v_s$ provides the twin baryon symmetry breaking scale. As $S$ doesn't couple to Higgs bosons at the 1-loop level, it's mass can be somewhat higher than that of the fields ${X,\tilde X}$.  We also assign specific gauge and global charges for $\tilde{X}$, $X$ and their $\mathbb{Z}_2$ partners so that each of them couples to one type of fermion bilinears and all quarks have their masses lifted. (See table \ref{tb:strongsolncharge}.)
\begin{table}
\begin{center}
\begin{tabular}[t]{ |c | c | c | c|}
  \hline
  Names & $\tilde{X}$ &X & S   \\ 
   \hline
 $[SU(3)_c\times U(1)_Y]_{B}$ & $\left(\bar{3}, \frac{1}{3}\right)$ & $(3, -\frac13)$ & $(1, 0)$ \\
   \hline
   $Q_{\text{bn}, {B}}$ & $-\frac13$& $-\frac23$&$-1$\\
   \hline
   $Q_{\text{bn}, {B}}-Q_L$ & $\frac23$ &$-\frac23$ & 0\\
   \hline
  \end{tabular}
  \caption{Charges of new scalar fields (in sector B) for the model of section \ref{sec:strong}.}\label{tb:strongsolncharge}
\end{center}
\end{table}
Twin symmetry requires  twin counterparts $\tilde{x}$, $x$ and $s$ with the same charge assignment by exchanging Twin $\leftrightarrow$ SM. We suppose that the scalar fields $\tilde{X}$ and $X$ have similar physical properties, and note  that they carry opposite $B-L$ charges. This motivates us to consider a $\mathbb{Z}_{2}$ symmetry exchanging $\tilde{X}\leftrightarrow X$ and $\tilde{x}\leftrightarrow x$ in the theory. While this is broken by the low energy theory, it helps motivate a more constrained potential. The scalar potential with respect to this symmetry is written as, 
\bea
\kappa_x\left[|\tilde{X}|^4+|X|^4+2\alpha_{x}|\tilde{X}|^2|X|^2-2\xi_s |S|^2(|\tilde{X}|^2+|X|^2)\right]+\kappa_b \left|X\tilde{X}-\mu_b S\right|^2+\mathbb{Z}_2^{\text{twin}}
\eea
where we take $0<\xi_s\ll 1$ and $|\mu_b| < \xi_s v_s$ so that the back reactions from $\tilde{X}$ and $X$ to $S$ are small.  Scalar quadratics $m_x^2 (|X|^2+|x|^2+|\tilde{X}|^2+|\tilde{x}|^2)$ may also appear in the tree level Lagrangian and will not affect the vacuum structure qualitatively as long as $0< m_x^2<\xi_s v_s^2$ is held.
We also require $|\alpha_x| <1$ and positively valued parameters $\kappa_{b, x}\sim \mathcal{O}(1)$ so that the potential is bounded from below. When all the dust has settled, we may choose the energetically favorable vacuum alignment of those scalars which saturates the lower bound of the potential,
\bea
\langle\tilde{X}\rangle = \bmtx 0\\0\\ v_{\tilde{X}}\emtx,\quad \langle X\rangle =\bmtx 0\\v_{X_2} \\ v_{X_3} \emtx,\quad \langle S\rangle \simeq v_s, \quad \langle \tilde{x}\rangle =\langle x\rangle = \langle s\rangle =\vec{0}\;,
\eea 
where $"\simeq"$ is equality up to corrections from back reactions. In the limit of $m_x=0$ and $|\alpha_x|\ll 1$ the components of VEVs are given as, 
\bea
 v_{\tilde{X}}\simeq\sqrt{\xi_s/(1+\alpha_x)}v_s\;,\quad v_{X_3}\simeq\sqrt{(1+\alpha_x)/\xi_s}\mu_b\;,\quad v_{X_2}=\sqrt{v_{\tilde{X}}^2-v_{X_3}^2}
\eea
Therefore in the Twin sector $[SU(3)_c\times U(1)_Y]_B$ breaks into $\widehat{U}(1)_X$. Since the twin baryon symmetry is spontaneously broken there is one Goldstone boson left uneaten. Such massless scalar will decouple from the thermal bath at this SSB scale ($T \sim$ few TeV), and yield a tiny contribution to $\neff$ at late times.  By contrast, both of the strong gauge and $U(1)_{\text{bn}}$ symmetries are intact in the visible sector so no proton decay appears in our model.
We write down the following Yukawa interactions
\bea
\mathcal{L}_Y& \supset  Q  \left(\hat{\lambda}_{L} \tilde{X} L+\hat{g}_{L}XQ\right)+U_R^c\left(\hat{\lambda}_{R}\tilde{X}^\dagger  E_R^c+2\hat{g}_{R}X^\dagger  D_R^c\right)+\;\mathbb{Z}_2^{\text{twin}}\\
&-\left( \tilde{x}^\dagger d_R^c+ \tilde{X}^\dagger D_R^c\right)\hat{\lambda}_n N+\left(H_A\ell+H_B L\right) \hat{Y}_{\nu}N
\eea
where $\hat{\lambda}_{L, R, n}$ and $\hat{g}_{L, R}$ are $3\times 3$ Yukawa coupling matrices and we can make the simplifying ansatz: $(\hat{\lambda}_{L, R, n})_{ij}=\lambda_{L,R, n}\delta_{ij}$ and $(\hat{g}_{L, R})_{ij}=g_{L,R}\delta_{ij}$ for minimal flavor violation (MFV). Noting that $q\ell= u e-\nu d$, so the visible leptoquark $\tilde{\phi}$ does not interact with $de$, and cannot mediate processes such as $K_L\rightarrow\mu e$ at tree level.  

\begin{figure}[t]
\centering
\includegraphics[width=0.6\textwidth]{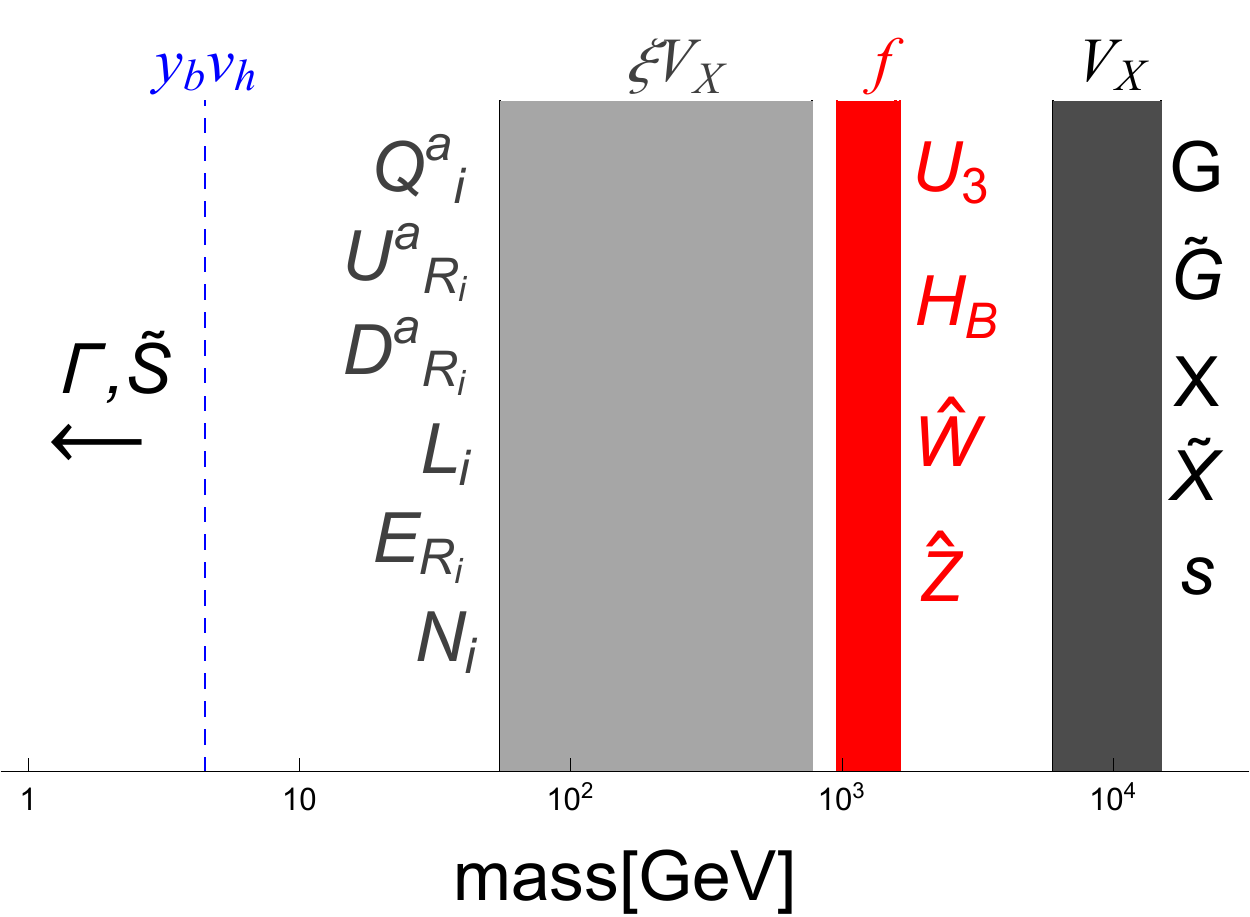}
\caption{A schematic of the spectrum of the theory with the field content in table \ref{tb:strongsolncharge}.}
\label{fig:strongspectrum}
\end{figure}

A schematic of the resulting spectrum is shown in Figure \ref{fig:strongspectrum}.
The top quark is still the heaviest twin fermion whose mass is mainly obtained from $H_B$, but mass spectrum for other twin particles is significantly lifted. For instance, taking $v_s=50$ TeV, $\xi_s=0.1$, $\alpha_x=0$, $\mu_b=3.6$ TeV and $m_x/\sqrt{2\kappa_x} = 5$ TeV, we will have $v_{\tilde{X}}=15$ TeV, $v_{X_3}=12$ TeV and $v_{X_2}=9$ TeV. Then except the massless twin photon and the neutral Goldstone boson, all the twin particles masses can be of the order of $100$ GeV if $|\lambda_{L, R, n}|\sim |g_{L,R}| \sim0.01$. Also, as these particles go non-relativistic and the sectors ultimately decouple, the twin photon decoupling temperature is around this energy scale. 
The massless SM neutrinos are combinations of dominantly $\nu_i$ and subdominantly $D_R$. Twin left-handed neutrinos get large masses from $v_{\tilde{X}}$ and are removed from low energy theory. For sterile neutrino masses  about $10^5$ MeV, we have an upper bound of the neutrino mixing angle $U_{\tau n}\approx Y_\nu v_h/(\lambda_n v_{\tilde{X}})< 10^{-2}$. Therefore $|Y_{\nu}| \leq 10^{-2}$ is allowed according to all the current bounds, which is a fairly weak constraint. 
Moreover, leptoquark and diquark masses are largely irrelevant to the hierarchy problem due to their smallness, this spectrum does not spoil the twin Higgs mechanism.\\

\subsection{Signals from Colliders and Flavor}
Both leptoquark and diquark scalars appear in various extensions of the SM and have been extensively studied e.g., \cite{Han:2009ya, Gogoladze:2010xd}. For instance, scalar quarks in supersymmetric models without R-parity may have either leptoquark-type or diquark-type Yukawa couplings. In grand unified theories (GUTs) like $SU(5)$, $SO(10)$ and $E_6$, leptoquarks or color triplet diquarks are also well motivated since they can be accommodated into matter representations.

In our strong solution model, these fields are present and needed to help lift the Twin spectrum. We expect couplings to all generations, although aside from giving masses, we do not have strong constraints on the sizes and couplings, although we typically take them to be small $\sim 0.01-0.1$. 

Leptoquarks and diquarks  are already very constrained. Generically, for diquarks with large Yukawas to light generations, there are constraints from dijets while with small Yukawas ($\lesssim 0.01$) the constraints come from QCD pair production and resulting 4-jet production.  

For  small Yukawa couplings, as is typically relevant here, searches for paired dijet resonances yield limits of $M_{DQ} \gtrsim 400 \gev$  ($36.7 {\rm fb^{-1}}$ and $35.9 {\rm fb^{-1}}$) \cite{Aaboud:2017nmi,Sirunyan:2018rlj}. For heavy flavor, this becomes somewhat stronger $\sim 600 \gev$. For larger Yukawa couplings, the lower bound on scalar diquark production from dijet analyses provided by CMS are  $ M_{DQ} \gtrsim 7.2\tev$ ($36 {\rm fb^{-1}}$)~\cite{Sirunyan:2018xlo}.  
 This is limited to situations where the resonance is narrower than the experimental resolution, which is generally true for our models. 
 
Constraints on leptoquarks  ($M_{LQ}$) can be quite strong. 
For us, typically, the most relevant bounds come from pair production. The precise limits depend on final states, which can be  $e_i e_i j j$ channel or $e_i\nu_i j j$ for our models. 
For the first generation leptoquarks, assuming the decay is entirely to $eejj$ channel, $M_{LQ}$ less than $\sim 1400\gev$ are excluded. For leptoquarks decaying to $e\nu jj$ channel alone, the lower bounds on $M_{LQ}$ is slightly weaker $\sim 1200\gev$~\cite{Sirunyan:2018btu,Aaboud:2019jcc}. Second generation LQ bounds are similar. For the third generation, leptoquarks decaying into $\tau$ and $b$ quarks has been presented in~\cite{Sirunyan:2018jdk,Aaboud:2019bye}. These limits tend to be somewhat weaker $\sim \tev$. 
 
 Precise LHC measurements of Drell-Yan kinematics constrain new physics~\cite{deBlas:2013qqa,Alves:2014cda,Farina:2016rws,Alioli:2017nzr}, as shown for t-channel leptoquark exchange in neutral current~\cite{Raj:2016aky} and charged current~\cite{Bansal:2018eha} events. For us, these limits are typically not relevant as the couplings are sufficiently small.

Recently, there have been claims of anomalous rates of B meson decay observed at BaBar, Belle, and LHCb. As a result, there has been significant interest in $\tev$ scale scalar leptoquarks~\cite{Bauer:2015knc}. In particular, the decay rate $\Gamma (\bar{B}\rightarrow D^{(*)} \tau\bar{\nu})$ is found to be about 30\% larger than predicted in the SM. A common origin for such excess is explained by the effects of the four-fermion operator $(c_L^\dagger \bar{\sigma}^\mu b_L)(\tau_L^\dagger\bar{\sigma}_\mu \nu_L)$ which can be produced from  leptoquark exchange. Good fits to the experimental data can be obtained with  scalar leptoquark models~\cite{Freytsis:2015qca}, although with only Minimal Flavor Violation there is tension with other experiments \cite{Bansal:2018nwp}. Considering non-MFV scenarios is beyond our scope here. We mention here this as a possible signature from our model and will leave the details for study elsewhere. 

\section{Conclusions}

The absence of LHC signals have increasingly pushed us to consider scenarios of new physics that are inaccessible to current searches. The natural directions for this are kinematical (the particles are too heavy to be seen or be seen easily) or sequestering (a sector with weak couplings to the SM). Models of neutral naturalness and the Twin Higgs, specifically, offer a candidate of this latter category.

These models suffer from cosmological tensions arising from the presence of new light states. A challenge is how to lift them without spoiling the $\ztwo$, and, moreover, to ensure that the entropy of the twin sector ultimately is small or can shift to the SM before BBN.

In this paper, we have attempted to address both of these, as the solutions go hand-in-hand. We have shown simple models of $\ztwo$ breaking, specifically where the Twin hypercharge is spontaneously broken by a $U(1)$ charged scalar, which lifts many Twin degrees of freedom, including the Twin photon and up to two neutrinos. But such a $\ztwo$ breaking does not address the cosmological issues. To that end, we have considered the role a Portalino can play to allow a further lifting of the degrees of freedom and a transfer of entropy to the SM.

Such models naturally yield mixings between the SM neutrino and Twin fermions. In order that these mixings be small, some limited additional matter must be present. This can arise on top of the model already discussed, in the form of an extra Higgs doublet, or by breaking color in the Twin sector during $\ztwo$ breaking, rather than hypercharge. These models then produce quite different collider signals - one predicting no new colored matter, but interesting new electroweak states, the other predicting diquarks and leptoquarks that may be detectable at the LHC or future colliders.

It is intriguing that as one searches for viable models of neutral naturalness, one finds they must naturally have additional matter. And, indeed, in going beyond the Standard Model with matter charged under the SM naturally provides new signals. In this fashion, we may find that neutral naturalness is not {\em completely} neutral after all.

\vskip 0.25in

\noindent{\bf Note added:} As this manuscript was in preparation, \cite{Batell:2019ptb} appeared which contains the $U(1)$ breaking model of section \ref{sec:u1break}.

\acknowledgments
The authors thank Matthew McCullough, Martin Schmaltz and Joshua Ruderman for helpful comments. NW is supported by the Simons Foundation and by the NSF under grant PHY-1620727.
\appendix
\section{Reheating}\label{appx}
We provide here the explicit discussion of the reheating scenario that reduces $\neff$ with portalinos from the first two families. In this approach, the hidden pions are supposed to be short-lived for the twin tauon is lighter than $m_\tau f/v_h$ due to the nonvanishing overlaps with other twin leptons via $\hat{Y}_\phi^{\mu\tau}$ or $\hat{Y}_\phi^{\tau e}$ terms. To be specific, without the assumption that $Y_{\phi_{1}}=Y_{\phi_2}=0$, we can rewrite the last term in e.q.~(\ref{LTP}) as $\mathcal{L}\supset M_\Psi \Psi_{L_r}\Psi_{L_z}$ with,
\bea\label{phiLL}
\Psi_{L_r}= s_{\omega_r} L_1-c_{\omega_r} L_2\;,\; \Psi_{L_z}= s_{\omega_z}\left(c_{\omega_r} L_1+s_{\omega_r} L_2\right)-c_{\omega_z}L_3\;,\; M_\Psi=Y_\phi v_\phi\;,
\eea
where $Y_\phi = \sqrt{|Y_{\phi_i}|^2}$ and we have defined two angles in the lepton flavor space $s_{\omega_z}=Y_{\phi_3}/Y_\phi$, $s_{\omega_r}=Y_{\phi_2}/\sqrt{|Y_{\phi_1}|^2+|Y_{\phi_2}|^2}$.\footnote{In general, there is another flavor angle $\omega_\theta$ which is used to parametrize the degenerate mass eigenstates in the $\Psi_{L_r}-\Psi_{L_z}$ plane. Since this degeneracy doesn't affect the low energy physics when $M_\phi\gg f$, we only take a specific value of $\omega_\theta$ in our discussion.}  Then we have the third independent linear combination of $L_i$ which is the normal vector to the $\Psi_{L_r}-\Psi_{L_z}$ plane and does not get mass from $v_\phi$,
\bea
\Psi_{L_0}= c_{\omega_z}(c_{\omega_r}L_1+s_{\omega_r}L_2)+s_{\omega_z}L_3\;, 
\eea
which becomes the mode with zero eigenvalue of $\hat{Y}_\phi$. 
In the mass basis, $L_3$ can be substituted by $s_{\omega_z}\Psi_{L_0}$ as massive fields $\Psi_{L_r}$ and $\Psi_{L_z}$ are decoupled from the low energy. Being an $SU(2)$ doublet it is convenient to clarify the two components of the field as $\Psi_{L_0}= (\Psi_{\hat{\nu}_3}, \Psi_{E_{3}})^T$ which specifies that $\Psi_{E_{3}}$ obtains Dirac mass $M_{E_3}=m_\tau s_{\omega_z} f/v_h$ with $E_{R_3}$ through the twin $\tau$ Yukawa interaction and $\Psi_{\hat{\nu}_3}$ remains massless. If $M_{E_3}$ is smaller than the mass of twin pions, $M_{\Pi}^2= m_{\pi}^2 f/v_h \simeq (0.38\; \text{GeV})^2$, the decay channel $\Pi^\pm\rightarrow \Psi_{\hat{\nu}_3}\Psi_{E_3}$ is kinematically allowed and  $\Psi_{\hat{\nu}_3}$ will be the final relativistic relic in the twin sector.

Notice that the massive portalinos $N_{1, 2}$ interact with both sectors, and in principle they can decay both visibly and in the Twin sector. In this work, we assume $\Psi_{L_r}$ and $\Psi_{L_z}$ are heavier than $N_{1, 2}$ so that portalinos only decay into SM sector. As the Universe cools down, the two sectors keep in thermal equilibrium via Higgs portal at the high temperature era and decouple at $T_D\sim 5$ GeV. In the SM regime, besides photons and gluons, we include the fermions with mass between neutrinos and tau, i.e. $g_{*A}^D=75.75$. In the hidden sector we have $\Psi_{L_0}$, $E_{R_3}$, gluons and three light quarks which gives us $g_{*B}^D=52.75$. As we have mentioned in sec.~\ref{primer}, portalinos only decay into SM sector via the channel $N_a\rightarrow\phi\rightarrow e_{R}^c\ell\ell$, 
\begin{figure}[t]
\centering
\includegraphics[width=0.5\textwidth]{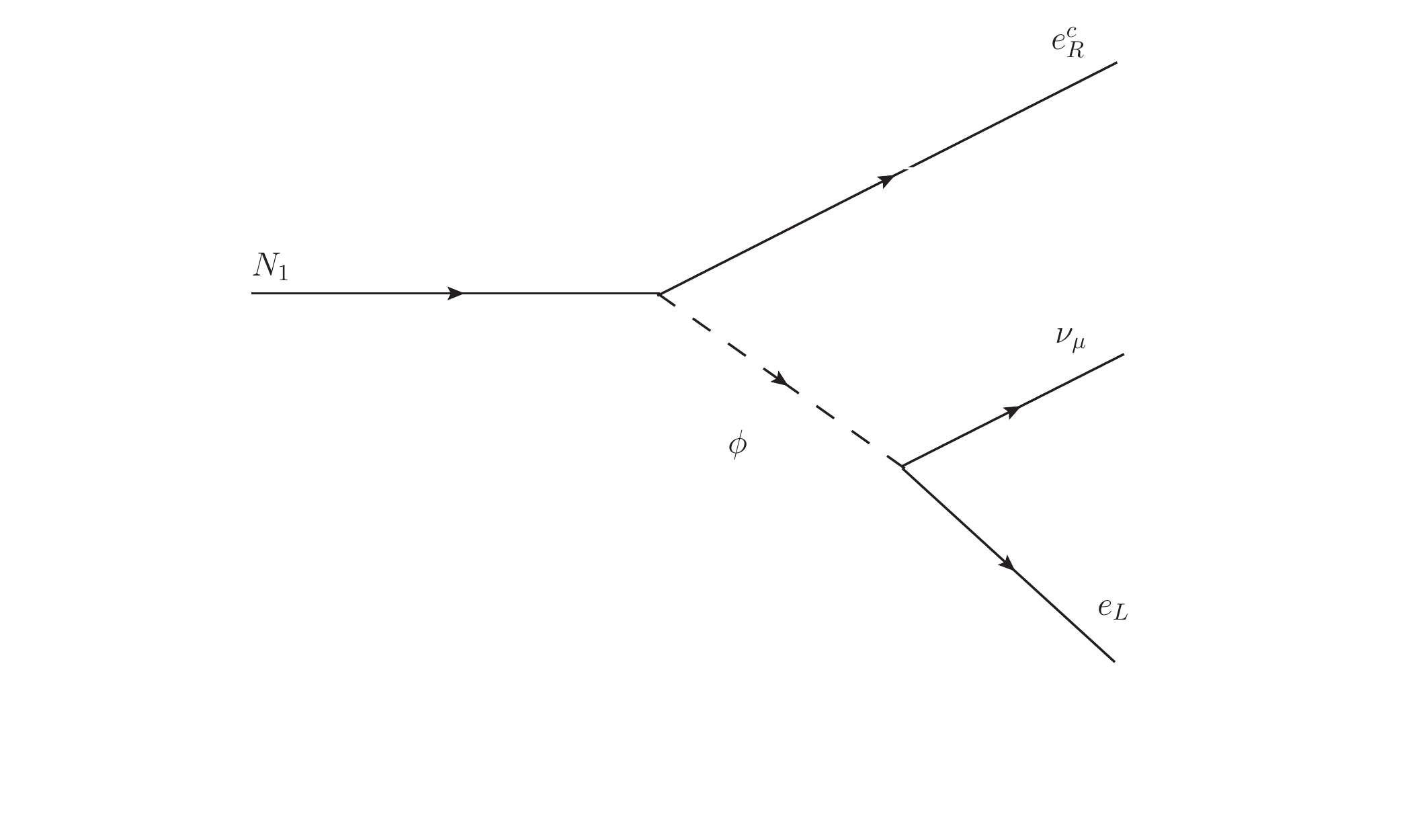}
\caption{Portalino decay into SM muon and neutrino}
\end{figure}

Then the visible sector will be significantly ``heated up'' and be in consistence with the viable cosmology if~\cite{Chacko:2016hvu}
1) $N_a$ go out of the thermal bath when they are still relativistic. 
2) Most of $N_a$ must be depleted before the SM temperature falls below an MeV.
3) Most of $N_a$ decays must occur after Twin-SM decoupling. 
The conditions above are equivalent to $H_{\nu d}<\Gamma_N < H_D/5$ and $m_N< T_0$, where $H_{\nu d}$ and $H_D$ are Hubble parameters when SM neutrino decoupling occurs and Twin-SM decoupling occurs respectively.  $m_N=Y_E v_\phi$ is the mass of $N_a$ which turns to freeze out at the temperature $T_0\sim [m_\phi^4/(Y_E^2 Y_\phi^2 M_{Pl})]^{1/3}$. On one hand, the $N_a$'s decay rate $\Gamma_N$ can be estimated as, 
\begin{equation}\label{gamma}
\Gamma_N=\frac{Y_E^2 Y_\phi^2}{m_{\phi}^4}\frac{m_N^5}{192\pi^3}
\end{equation}
On the other hand, assuming that by the time of reheating $t_R\sim\Gamma_N^{-1}$, $N_a$'s energy density dominates that of the SM before decays, i.e. $\rho_N =3 M_{Pl}^2 \Gamma_N^2 $. 
Suppose those energy immediately transfers into SM radiation, $\rho_R^A=\rho_N$, we will have the ratio of energies between Twin and SM sectors, 
\bea\label{YE2}
\frac{\rho_B}{\rho_A}=C_R \left(\frac{\Gamma_N^2 M_{Pl}^2}{g_{* B, R} m_N^4}\right)^{1/3}\leq \bar{R}_N
\eea
where $C_R=(\pi^2/90)[4\pi^2 g_{*0} g_{* B, D}/(\zeta(3) g_{* N} g_{* D})]^{4/3}$ and we denote $\bar{R}_N$ as the upper bound of dark radiation ratio that is capable of realizing this scenario successfully. Plugging in the expression of $\Gamma_N$ in E.q.~(\ref{gamma}), 
we get the constraint on $Y_E$,
\bea
\left(\frac{64\pi^4 T_{\nu d}^2}{Y_\phi^2 M_{Pl} v_\phi}\sqrt{\frac{g_{*A, \nu d}}{10}}\right)^{1/7}<Y_E<\left(\frac{192\pi^3v_\phi}{Y_\phi^2 M_{Pl}}\sqrt{\frac{g_{* B, R}\bar{R}_N^3}{C_R^3}}\right)^{1/5}
\eea
For larger $v_\phi$, the range of $Y_E$ is broader. For instance, if taking: $T_{\nu d}= 1$ MeV, $v_\phi=10$ TeV, $Y_\phi=0.01$, $\bar{R}_N=0.1$, we will have $9.4\times 10^{-4}<Y_E<2.7\times 10^{-3}$.
\bibliography{twinportalino}
\bibliographystyle{JHEP}
\end{document}